\documentclass[aps, apl, a4paper, reprint]{revtex4-1}
\usepackage{graphicx}
\graphicspath{./figures/}
\DeclareGraphicsExtensions{.pdf, .png}
\usepackage{overpic}
\usepackage{tikz}
\usepackage{siunitx}
\usepackage{color}
\usepackage{placeins} % for FloatBarrier
\usepackage{import}

\usepackage{epstopdf}
\PrependGraphicsExtensions{.tif, .tiff}

\begin{document}

\title{Understanding the Plasmonics of Nanostructured Atomic Force Microscopy Tips}
\author{A. \surname{Sanders}}
\author{R.W. \surname{Bowman}}
\author{L. \surname{Zhang}}
\author{V. \surname{Turek}}
\author{D.O. \surname{Sigle}}
\author{A. \surname{Lombardi}}
\author{L. \surname{Weller}}
\author{J.J. \surname{Baumberg}}
\affiliation{Nanophotonics Centre, Department of Physics, Cavendish Laboratory, Cambridge, CB3 0HE}
\email[Email: ]{jjb12@cam.ac.uk}
\homepage[Site: ]{www.np.phy.cam.ac.uk}
\date{\today}


\begin{abstract}
Structured metallic tips are increasingly important for optical spectroscopies such as tip-enhanced Raman spectroscopy (TERS), with plasmonic resonances frequently cited as a mechanism for electric field enhancement.  We probe the local optical response of sharp and spherical-tipped atomic force microscopy (AFM) tips using a scanning hyperspectral imaging technique to identify plasmonic behaviour. Localised surface plasmon resonances which radiatively couple with far-field light are found only for spherical AFM tips, with little response for sharp AFM tips, in agreement with numerical simulations of the near-field response. The precise tip geometry is thus crucial for plasmon-enhanced spectroscopies, and the typical sharp cones are not preferred.
%These results demonstrate the benefits of nanostructuring tips and the need to understand and characterise the plasmons present in a nanostructure prior to use as optical antennae / near-field enhancers.
\end{abstract}

\maketitle

% most simulations just use a small cone, resonances

% mention nanoscale localisation of light
Within the last decade nano-optics has benefited from the advent of metallic tip-based near-field enhancement techniques such as TERS and scanning near-field microscopy (SNOM), leading to successes in single molecule detection \cite{zhang2013} and spatial mapping of chemical species \cite{pettinger2012}. Despite their high spatial resolution and scanning capabilities, there remains confusion about the plasmonic response of metallic tips.
%Their high resolution and spatial scanning capability mean these techniques are widely considered to become the successors of static techniques like SERS. However, for this to be the case, the near- and far-field optical response of metallic tips must be well understood.
Tip systems built on AFM probes can exhibit electric field enhancements close to 100 at the apex (Raman enhancements up to \num{e8}) \cite{pettinger2012}, due to a combination of plasmonic localisation and a non-resonant lightning rod effect. %The majority of enhancement in sharp tips is a result of the lightning rod effect
%Plasmon excitation works to overcome the 3--4 orders of magnitude coupling efficiency loss between diffraction-limited light and the nanometre scale \cite{berweger2010}. A metallic nanostructure supporting plasmons can therefore act as an optical antenna, mediating energy transfer between far-field radiation and the near-field \cite{novotny2006, novotny2011}.
The factors determining a tip's ability to enhance the near-field include the experimental excitation/collection geometry, tip sharpness, surface metal morphology, and constituent material.

Despite large measured near-field enhancements, the standard sharp AFM tip geometry does not support radiative plasmons. 
%{\color{red}The majority of tip-induced field enhancement likely results from a non-resonant lightning rod effect at the sharp tip apex.} 
The extended ($\sim$\SI{20}{\micro\metre}) size and single curved metal-dielectric interface of an AFM tip supports only weakly confined localised surface plasmons (LSPs) \cite{zhang2009} and propagating surface plasmon polaritons (SPPs), which may be localised by adiabatic nanofocussing \cite{stockman2004, pile2006, berweger2010, lee2011, berweger2012, lindquist2013}. Lack of a dipole moment means that neither LSPs or SPPs strongly couple with radiative light in the same manner as multipolar plasmons in sub-wavelength nanoparticles \cite{zhang2009}. For this reason, the tip near-field is often excited with evanescent waves \cite{hamann1998} or via nanofabricated gratings \cite{berweger2010} to access the optically-dark SPPs, with resonant scattering of evanescent waves \cite{neacsu2005, mehtani2006, barrios2009}, resonances in the TERS background \cite{pettinger2007, pettinger2009} and depolarised scattering images \cite{mino2014} providing evidence for localised plasmon excitation. For Au tips such plasmon resonances are typically found between 600--800\,nm.

\begin{figure}
\centering
\includegraphics[width=0.9\columnwidth]{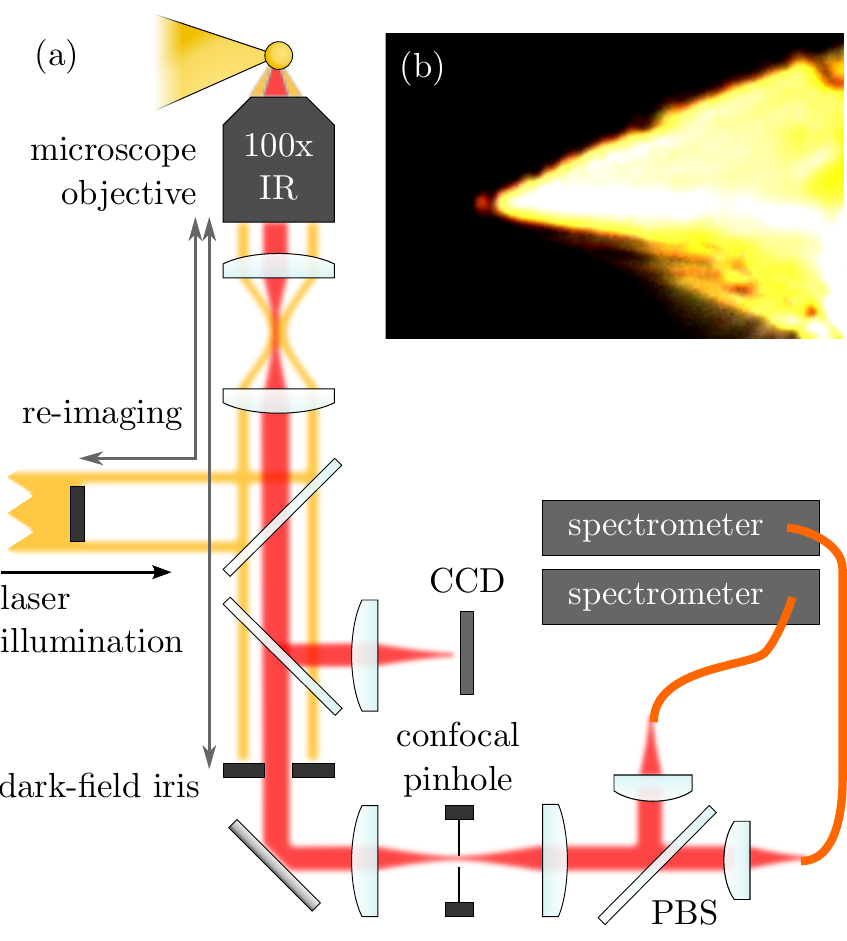}
\caption{(a) Hyperspectral imaging with supercontinuum laser focussed onto the tip apex for imaging, and the tip is raster scanned across the beam with scattering spectra of both polarisations acquired at each position. (b) Ball-tip imaged in dark-field microscopy.
%A central stop in the illumination beam and an iris in the detection path are re-imaged onto the objective's back aperture to eliminate reflection and measure only scattered light.  A polarising beamsplitter (PBS) directs the axially and longitudinally polarised light to two different fibre-coupled spectrometers.  The inset shows a spherical Au AFM tip as viewed in a dark-field optical microscope.
}
\label{fig:simple_optics_layout}
\end{figure}

%Theory predicts that whilst shorter, truncated, tip geometries (nano-cones, ellipsoids or pyramids) show visible-NIR multipolar plasmons \cite{roth2006, goncharenko2006, schafer2013, cherukulappurath2013}, these redshift and diminish with increasing tip length, leaving only the smooth response of the lightning rod effect, increasing towards the IR \cite{zhang2009, huber2014}. The standard, sharp metallic tip geometry therefore makes for a poor \emph{plasmonic} optical antenna unless the optically-dark SPP modes can be accessed. For this reason, the tip near-field is often illuminated with evanescent waves to enable plasmon excitation \cite{}, with resonant scattering of evanescent waves \cite{neacsu2005, mehtani2006, barrios2009}, resonances in the TERS background \cite{pettinger2007, pettinger2009} and depolarised scattering images \cite{mino2014} providing evidence for localised plasmon excitation. For Au tips such plasmon resonances are typically found between 600--800\,nm.

Improvements in enhancement are often found in roughened tips with grains acting as individual nano-antennae for more confined LSPs, however this approach lacks reproducibility \cite{mino2014}.
In recent years controlled nanostructuring of the tip apex with a distinct sub-wavelength-size metallic feature has been explored in order to engineer and tune a plasmonic optical antenna precisely at the apex and better incorporate more localised multipolar plasmons \cite{hayazawa2001, bailo2008, hayazawa2012, umakoshi2012, kharintsev2013, mino2014}. Etching \cite{uebel2013, kharintsev2013}, focussed-ion-beam machining \cite{weber2010, fleischer2011, maouli2015}, selective deposition \cite{zou2009}, nanoparticle pickup \cite{denisyuk2012}, nanostructure grafting \cite{huth2013} and electrochemical deposition \cite{sanders2014} have all been successfully used to nanostructure optical antenna tips.

Scattering resonances in the visible-NIR spectrum have been directly measured on a subset of these \cite{zou2009, sanders2014, maouli2015} while other reports use improvements in the field enhancement as a measurement of antenna quality \cite{umakoshi2012, huth2013, kharintsev2013}. In such cases the field enhancement has been attributed to give improvements by an order of magnitude through plasmon excitation \cite{weber2010, fleischer2011, umakoshi2012, sanders2014}.

The simplest geometry for a tip apex is a spherical nanoparticle (NP), giving LSPs similar to those in an isolated spherical metallic nanoparticle. In this paper we demonstrate an effective method for characterising the radiative plasmon modes of a tip and clearly show the benefits of utilising spherically nanostructured tips as near-field enhancers.

%\section{Experimental}

% Introduce the experimental part of the studies
The optical properties of AFM tips are studied using a custom-built confocal microscope with a supercontinuum laser source for dark-field scattering spectroscopy (Fig.~\ref{fig:simple_optics_layout}). Both illumination and collection share the optical axis of a 0.8\,NA IR objective. Supercontinuum laser light is filtered into a ring and incident on a tip at 0.6--0.8\,NA while light scattered by the tip is confocally collected from the central laser focus using an iris to restrict the collection NA below 0.6. Broadband polarising beamsplitters are used to simultaneously measure spectra which are linearly polarised both along the tip axis (axial) and perpendicular to the tip axis (transverse).

% Introduction to hyperspectral imaging in the experimental
A scanning hyperspectral imaging technique is applied to determine the local optical response at the tip apex. Tips are raster scanned under the laser spot and the dark field scattering from the confocal sampling volume measured at each point, forming a hyperspectral data cube. Images are formed at each wavelength contained in the cube, with each image pixel digitised into 1044 wavelengths between 400--1200\,nm. Measured spectra are normalised to a spectrum of flat metal of the same material to show only structural effects. Image slices at individual wavelengths or wavelength bands are then readily constructed to display localised spectral features. 
Fast image acquisition is made possible by the high brightness supercontinuum laser source (100 $\mu$W.$\mu$m$^{-2}$) and cooled benchtop spectrometers, enabling 10\,ms integration times (5 mins per image). 
%Image acquisition takes around 5 minutes for $100\times100$ pixels.
%The focal intensity is not sufficiently high enough to damage the \SI{50}{nm} metallic tip coatings.
%The illumination and collection configuration remains fixed between different samples to better compare images.
Within plasmonics, this approach to hyperspectral imaging has been used to identify distributed plasmon modes in aggregated AuNP colloids \cite{herrmann2013} and to image SPPs \cite{bashevoy2007hyperspectral} but has yet to be applied to tips. By using this technique, radiative plasmons can be spatially identified with a resolution around 250\,nm.

%\section{Results} 

\begin{figure}[t]
\centering
\includegraphics[width=0.98\columnwidth]{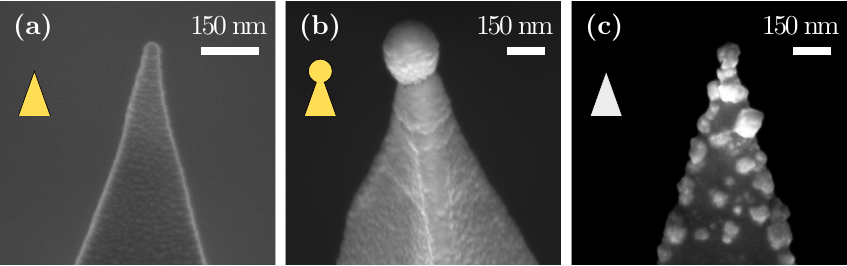}
\caption{SEM images of (a) sharp Au AFM tip, (b) Au-coated spherical AFM tip (Nanotools), and (c) electrochemically-deposited AuNP-on-Pt AFM tip.}
\label{fig:tip_sems}
\end{figure}

To investigate the radiative plasmonic properties of nanostructured tips, hyperspectral images are taken of both standard (sharp) and spherical-tipped Au AFM tips. Spherical tips are either 300\,nm diameter, 50\,nm Au-coated NanoTools B150 AFM probes or electrochemically-deposited AuNP-on-Pt AFM probes, fabricated in-house \cite{sanders2014} (shown in Fig.~\ref{fig:tip_sems}). Fabricated tips are pre-treated where possible prior to use with ambient air plasma and/or piranha solution to remove organic surface residue and, in some cases, smooth out surface roughness.

%\onecolumngrid
%\begin{figure*}
\begin{figure}[t]
\centering
\includegraphics[width=0.9\columnwidth]{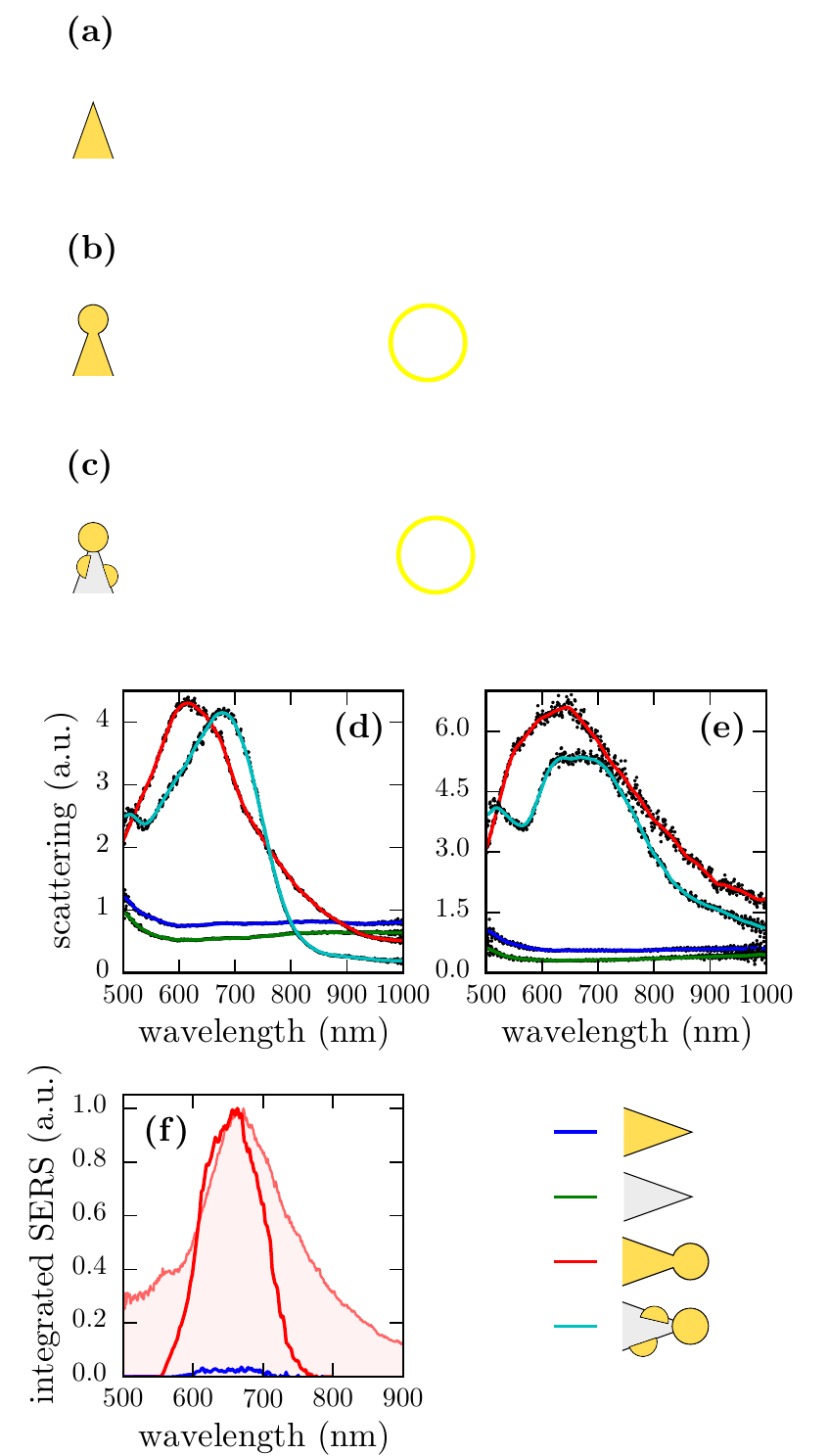}
\caption{
Hyperspectral images of (a) sharp Au tip, (b) Au-coated spherical tip (Nanotools), and (c) electrochemically-deposited AuNP-on-Pt tip. Collected light is polarised along tip axis, colour maps all have same normalisation. Scale bar is $600\,$nm.
%Resonant scattering from spherical apex is clearly seen in hyperspectral images between 600--700\,nm.
(d,e) Scattering spectra of both sharp and spherical metal tips, extracted from hyperspectral images around the apex region, in (d) axial and (e) transverse polarisations.
(f) Integrated SERS background from sharp and spherical Au tips. Scattering spectrum of spherical Au tip apex shown shaded.
%  probe the internal tip near-field and confirm plasmon excitation. The background spectrum is the 
}
\label{fig:hyperspectral_images}
\end{figure}
%\twocolumngrid

%\onecolumngrid
\begin{figure*}
\centering
\includegraphics{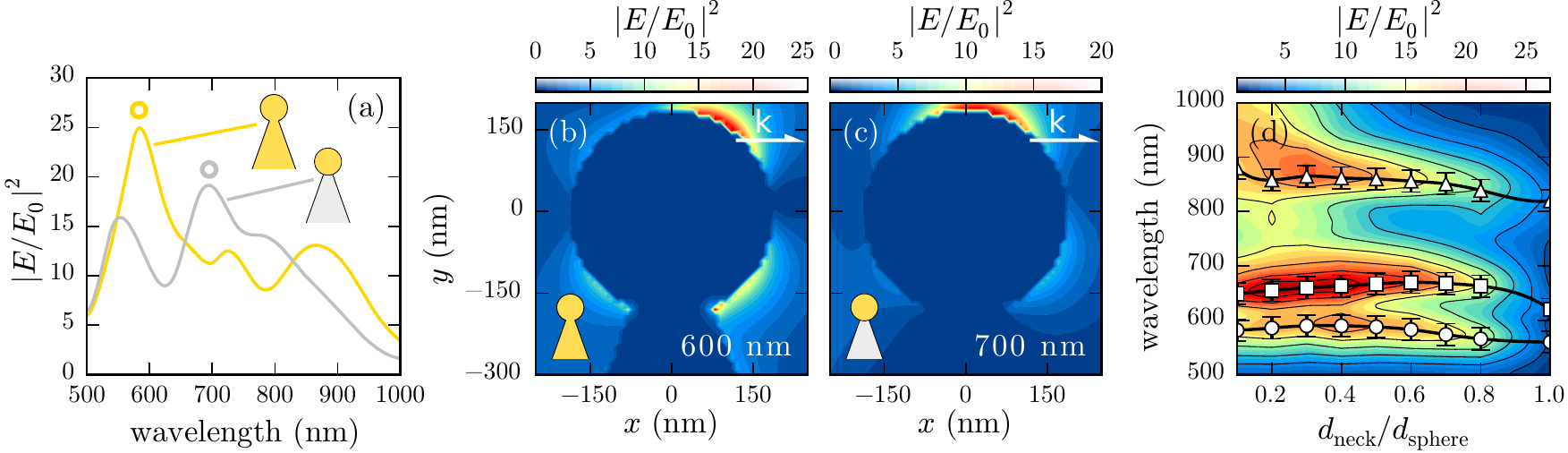}
\vspace{-5pt}
\caption{(a) Numerically simulated near-field apex spectra of spherical Au and AuNP-on-Pt tips with (b,c) near-field maps of the main resonance in each, as highlighted by circles in (a). Simulated tips have a 300\,nm spherical radii, 120\,nm neck widths, \SI{20}{\degree} opening angles and \SI{1.88}{\micro\metre} lengths to best match typical experimental tip geometries and avoid truncation artefacts. Tips are illuminated by plane waves orientated along the tip axis. (d) Interpolated field enhancement map with superimposed resonant wavelengths, as the neck width varies from a spherical to a sharp tip. Tips have a 250\,nm apex diameter, \SI{1.88}{\micro\metre} length, and \SI{10}{\degree} opening angle.}
\label{fig:numerical_simulations}
\end{figure*}
%\twocolumngrid

Comparisons between spherical- and sharp-tipped Au probes using hyperspectral image slices (Fig.~\ref{fig:hyperspectral_images}) shows that spherical tips exhibit a characteristic red (600--700\,nm) scatter, separated from the bulk tip. No similar localised scattering is seen in the visible spectrum with sharp Au tips, which have a ten-fold weaker optical response and appear similar to non-plasmonic Pt tips.  This delocalised apex scatter can also be directly seen in dark-field microscopy images (Fig.~\ref{fig:simple_optics_layout}b).
The AuNP-on-Pt structure behaves very similarly to the Au-coated spherical tip (which has diamond-like-carbon inside), likely because the 50\,nm coating thickness is greater than the skin depth \cite{stockman2011, huber2014}. 
As we show below, differences in plasmon resonances arise due to the Au-Pt and Au-Au neck boundaries.

%\begin{figure}[t]
%\centering
%\includegraphics{tip_spectra}
%\caption{
%Apex scattering spectra of sharp and spherical metal tips extracted from the hyperspectral images by integrating pixels around the apex region in both the axial (a) and transverse (b) polarisations.
%%A clear resonance between 600--700\,nm is observed with spherical tips in both polarisations whereas sharp metallic tips show comparatively flat spectra.
%Integrated broadband tuneable SERS background measurements (c) of both a sharp and spherical Au tip probe the internal tip near-field and confirm plasmon excitation. The background spectrum is the scattering spectrum of the spherical Au tip apex.
%}
%\label{fig:apex_spectra}
%\end{figure}

Integrating spectra around each tip better shows the 600--700\,nm scattering resonance from spherical Au tips (Fig.~\ref{fig:hyperspectral_images}d,e), which are reliably present in all spherical-tipped AFM probes, both vacuum-processed and electrochemically deposited. We attribute these to localised surface plasmon excitation, while electron microscopy confirms this resonance correlates only with spherical Au tip shapes.  The response of sharp Au tips shows no similar plasmonic features, while the slow rise in scattering towards the NIR is consistent with lightning rod scattering \cite{zhang2009}.

Broadband tuneable SERS measurements \cite{lombardi2016} confirm that the optical scattering resonance seen in spherical Au tips is indeed caused by radiative plasmon excitation. The trapped plasmon fields enhance optical processes on the surface such as surface-enhanced Raman scattering (SERS) and here we use the SERS background \cite{lombardi2016,hugall2015} as a reporter of the plasmonic near-field strength.
SERS background spectra are integrated across a range of excitation wavelengths between 500 and 700\,nm, spaced 10\,nm apart, to extract any scattering resonances. The resulting spectrum (Fig.~\ref{fig:hyperspectral_images}f) shows a distinct peak around the spherical Au tip scattering resonance, while no such resonance is seen for sharp Au tips.
Further confirmation stems from direct observation of plasmon coupling between spherical tips, as has been previously reported \cite{savage2012}.

%\section{Discussion}

Plasmon resonances in spherical AuNP tips correspond to \emph{radiative} antenna-like modes, similar to those in plasmonic nanoparticles, that efficiently couple far-field light into strong collective free electron oscillations without the need for SPP momentum matching. As with nanoparticles, the signature of these plasmons is an optical resonance indicating their large dipole moment (Fig.~\ref{fig:hyperspectral_images}d). Such radiative plasmons only form if multipolar surface charge oscillations are supported, requiring a structure with multiple metal-dielectric interfaces. Since spherical metallic tips possess a neck behind the tip, they can support NP plasmonics. Sharp tips do not have this back surface, hence cannot support radiative plasmon resonances, although the single metal-dielectric surface supports launching of evanescent SPPs and a strong lightning rod component.

%Partial loss of the neck introduces a conductive pathway in spherical tips, modifies surface charge restoring forces, and provides a secondary surface for self-interaction. It is therefore difficult to analytically describe spherical tips. Numerical simulations of the near-field around spherical tips, computed using BEMAX, are instead employed to better understand their response.

Simulated near-field spectra (using the boundary element method) around the apex of 300\,nm spherical Au and AuNP-on-Pt tips with 120\,nm neck diameters ($d_{\mathrm{neck}}=0.4d_{\mathrm{sphere}}$) are shown in Fig.~\ref{fig:numerical_simulations}a. Tips are simulated with a length of \SI{1.88}{\micro\metre} to avoid truncation artefacts which are commonly seen in tip simulations and erroneously suggest plasmonic performance even in sharp tips. Strong modes appear along the tip axis for all spherical tips between 550--700\,nm, as in experiments with peak wavelengths that match our hyperspectral results.
Near-field maps corresponding to the main resonance in each tip (Fig.~\ref{fig:numerical_simulations}b,c) show dipole-like resonances with the neck spatially splitting the underside of each mode, mixing it with quadrupolar modes and shifting it towards the blue.

% show a more quadrupole-like plasmon in the spherical Au tip with a weaker dipole-like resonance occurring above 700\,nm. Visible frequency quadrupolar modes are often observed in larger Au nanoparticles once the dipolar resonance shifts into the NIR. A similarly structured quadrupole plasmon could be expected in 300\,nm spherical Au tips between 550--600\,nm. The neck geometry may also inhibit a dipolar plasmon polarisation and reduce its confinement, thus favouring a quadrupolar distribution.

%Similar but more blueshifted plasmons are found in the AuNP-on-Pt tip, with a dominant 700\,nm dipole-like resonance and a 550\,nm quadrupole-like resonance.
%Electromagnetic coupling between a Au and a Pt surface is weaker than the interaction between two Au surfaces \cite{ren2004} and the additional non-plasmonic interface aids confinement, hence plasmons in the spherical Au apex are less redshifted when attached to a Pt tip. As a result, the dipole-like mode exists nearer to the visible and radiates more strongly than the quadrupole-like mode.

In order to directly compare the \emph{plasmonic} behaviour of spherical and sharp Au tips independent of lightning rod contributions, the neck width is incrementally increased. This allows us to study structures which smoothly transition from a nanoparticle attached to the apex of a sharp Au tip, into a rounded tip geometry, without the apex radius ever changing. The field enhancement and peak positions extracted from this morphology transition (Fig.~\ref{fig:numerical_simulations}d) show resonances  insensitive to the neck width until $d_{\mathrm{neck}}>0.8d_{\mathrm{sphere}}$, explaining the robustness of observed spherical tip plasmons between different tip morphologies. However a steady decrease in the field enhancement is observed once $d_{\mathrm{neck}}>0.4d_{\mathrm{sphere}}$, decreasing faster once $d_{\mathrm{neck}}>0.8d_{\mathrm{sphere}}$. This supports the claim that sharp tips cannot sustain antenna-like plasmons and that the majority of enhancement is from lightning rod effects. We note that the lateral spatial localisation of the field approaches $0.3 d_{\mathrm{sphere}}$ independent of this neck diameter.

%\subsection{Implications}
% tense for this part onwards?

These results demonstrate the importance of considering which plasmons might exist in a particular experiment and nanostructure geometry, and that it is vital to characterise nanostructures prior to their application.
Apex nanostructuring can controllably introduce radiative plasmons into the tip geometry, lifting the evanescent illumination restriction of sharp tips and permitting use of a wider range of microscope configurations. While the lightning rod effect will always contribute to the field enhancement and favour sharp tips, exploiting resonant plasmonic enhancement in a carefully optimised spherical tip can further improve the near-field enhancement. The spherical tip geometry and materials shown here are optimised for use with the typically-used 633\,nm laser wavelengths. 

Demonstrated interactions between spherical tip plasmons \cite{savage2012} also suggests coupling with an image charge in a planar surface is possible and could be used in nanometric tip-surface gaps to further localise the field on resonance with near infrared lasers. Exploiting radiative tip plasmons in this manner bridges the gap between SERS and conventional TERS, forming a spatially-mappable version of the highly successful nanoparticle-on-mirror geometry \cite{mertens2013, taylor2014}. These systems repeatedly produce Raman enhancements of up to \num{e7} with nanometric mode volumes, much like tips, and demonstrate that plasmonic gaps can exhibit comparatively large field enhancements without relying only on the lightning rod effect.

Secondly, without prior knowledge of the tip-system spectral response it is difficult to properly interpret any measurements, such as TERS spectra. Improved tip characterisation is crucial to understanding variations in TERS spectra. Standard, wide-field microscopy/spectroscopy is not a particularly effective tool for optically characterising tips. Instead, confocal hyperspectral imaging provides a viable method for mapping the local scattering response while broadband tuneable SERS offers a unique way of optically characterising the near-field. Incorporating these techniques into existing microscopes is relatively simple and will greatly improve the reliability of tip-based near-field microscopy.

%\section{Conclusions}

%In conclusion, we have shown that localised surface plasmons can be optically excited and measured in spherically-nanostructured tips using scanning-confocal hyperspectral imaging in a supercontinuum dark-field microscope. Sharp tips show no such radiative plasmonic behaviour. This technique enables plasmon-dependent applications, such as TERS, to pre-screen tips prior to use to better improve reliability and reproducibility. The plasmons present in a system determine what plasmonic phenomena are able to be experimentally observed, hence spherical tips can be used to dynamically investigate plasmonics.

\begin{acknowledgments}
The authors thank EPSRC grants EP/G060649/1 and EP/L027151/1, and ERC grant LINASS 320503 for funding and NanoTools for their services providing Au-coated spherical AFM tips.  RWB thanks Queens' College and the Royal Commission for the Exhibition of 1851 for financial support.
\end{acknowledgments}

\bibliography{citations}

\end{document}